\documentstyle[aaspp4,12pt,epsf]{article} 

\newcommand{\xte}{{\sl RXTE}}
\newcommand{\psec}{s$^{-1}$}
\def\pcm#1{cm$^{-#1}$}                
\newcommand{\grs}{GRS~1915$+$105}

\begin{document}

\title{Variable-Frequency QPOs from the Galactic Microquasar GRS~1915+105}

\author{Craig B. Markwardt\altaffilmark{1,2}, Jean H. Swank\altaffilmark{1},
        Ronald E. Taam\altaffilmark{3}}

\altaffiltext{1}{NASA/Goddard Space Flight Center, Code 662, Greenbelt, MD 20771; craigm@lheamail.gsfc.nasa.gov (CBM); swank@pcasun1.gsfc.nasa.gov (JHS)}
\altaffiltext{2}{National Research Council Resident Associate}
\altaffiltext{3}{Department of Physics and Astronomy, Northwestern University, Evanston, IL 60208; taam@ossenu.astro.nwu.edu}

\begin{abstract}
We show that the galactic microquasar \grs\ exhibits quasi-periodic
oscillations (QPOs) whose frequency varies continuously from 1--15~Hz,
during spectrally hard dips when the source is in a flaring state.  We
report here analyses of simultaneous energy spectra and power density
spectra at 4~s intervals.  The energy spectrum is well fit at each
time step by an optically thick accretion disk plus power law model,
while the power density spectrum consists of a varying red noise
component plus the variable frequency QPO.  The features of both
spectra are strongly correlated with one another.  The 1--15~Hz QPOs
appear when the power law component becomes hard and intense, and
themselves have an energy spectrum consistent with the power law
component (with root mean square amplitudes as high as 10\%).  The
frequency of the oscillations, however, is most strikingly correlated
with the parameters of the {\it thermal disk} component.  The tightest
correlation is between QPO frequency and the disk X-ray flux.  This fact
indicates that the properties of the QPO are not determined by solely
a disk or solely a corona.
\end{abstract}

\section{Introduction}

The galactic source \grs\ was first discovered in 1992 as an X-ray
transient by GRANAT/Watch (Castro-Tirado, Brandt \& Lund 1992), and
subsequent radio observations showed that it is a source of
superluminal jet-like outflows (Mirabel \& Rodriguez 1994).  \grs\ can
be quite luminous in X-rays ($L_x \ge 10^{39}$~erg~\psec), above the
Eddington limit for neutron stars (Greiner, Morgan \& Remillard
\markcite{gmr96}1996).  It also exhibits a rich set of variability in
X-rays (Morgan, Remillard \& Greiner 1997, hereafter MGR97).  Much
interest has focussed on a weak 67~Hz quasi-periodic oscillation (QPO)
feature that is sometimes present.  These oscillations have been
interpreted as the signature of oscillating material at or near the
radius of marginally stable orbits (MGR97, Nowak et al 1997) and
perhaps Lense-Thirring precession due to a maximally rotating black
hole (Cui, Zhang, Chen \markcite{czc98}1998).  On the occasions when
it is detected, the frequency of the 67~Hz QPO does not appear to vary
with total source intensity (MGR97).  \grs\ exhibits lower frequency
oscillatory behavior as well, including a 1--15~Hz variable frequency
QPO, a lower frequency complex of QPO modes (below 0.1~Hz), and a
stochastic low frequency noise component.  In this Letter we
investigate the properties of the 1--15~Hz QPO, further concentrating
on how the QPO frequency depends on the spectral parameters of the
source.

Chen, Swank \& Taam (1997, hereafter CST97) have noted that the
frequency of the 1--15~Hz QPO generally depends on the total source
intensity, but the dependence between the two is neither simple nor
unique (see also Chen, Taam \& Swank 1998).  The X-ray flux of \grs\
can vary by a factor of five or more, and the count rate from \grs\ is
high enough that temporally resolved spectroscopy is possible over
intervals of several seconds.  We exploit that fact here for an
observation where both the intensity and QPO frequency cover a wide
range.  We find that the spectrum is well represented by a combination
of optically thick thermal and power law components.  The
lower-frequency QPOs clearly appear only when \grs\ is in a spectrally
hard state.  Remarkably, however, the frequency of the QPO is very
strongly dependent on the fitted thermal flux.

In this paper we present evidence of the strong correlations between
the spectral and temporal properties.  Section~\ref{Sobs} describes
the observational techniques and results.  Section~\ref{Scorr} shows
the correlation first between the energy and Fourier spectra, and
second, between the QPO frequency and the fitted fluxes of the
individual spectral components.  In Section~\ref{Sdisc} we discuss the
results in light of current accretion disk models.  

\section{Observations and Analysis \label{Sobs}}
\begin{figure}[tbp]   
\centerline{\epsfxsize=4.3in{\epsfbox{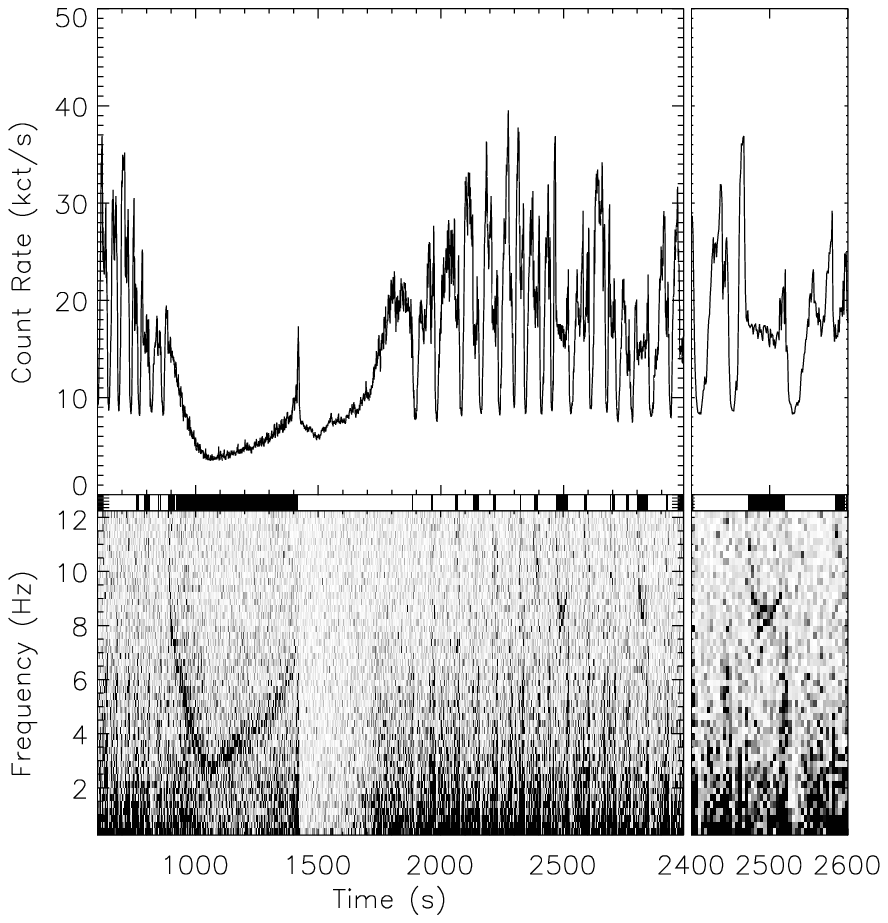}}} 
\caption{ 
(top) RXTE PCA light curve of \grs\ (for 4 PCU detectors). (bottom)
``Dynamical'' power density spectrum over the same time range, taken
at 4~s intervals.  The gray scale is the linear Leahy power density
from 0 (white) to 30 (black; significant at $\sim 5\sigma$ level).
The center band is black when a QPO is expected based on the energy
spectrum (see Table~\ref{Tstate}).\label{Flcqpo}}
\end{figure}

\grs\ has been the target of nearly 250 observations by \xte\ during
its mission between early 1996 to the present.  In this Letter we
present \xte\ spectral and timing results from observations of \grs\
on 1997~September~09.  Our analysis of observations from other days
indicate that this day is representative of a substantial fraction of
the source behavior, and that the source is variable enough to span a
large range in source intensity and QPO frequency.  A portion of the
light curve from the Proportional Counter Array (PCA) is shown in the
top panel of Figure~\ref{Flcqpo}.  On this particular day, the source
exhibits oscillatory ``flaring'' on time scales of $\sim$100~s,
followed by extreme dips which last several hundreds of seconds.  It
is these extreme dips which appear to be associated with the
disappearance of the inner accretion disk (Belloni et al
\markcite{b&97}1997) and simultaneous outbursts observed in the
infrared and radio (Mirabel et al \markcite{m&98}1998).  We find three
distinct states which \grs\ traverses repeatedly.

To resolve the rapid variations it is advantageous to measure spectral
changes on the shortest feasible time interval.  PCA data is collected
by the EDS in several concurrently running acquisition modes; each
mode can have differing temporal resolution and spectral coverage.
During this observation the ``binned'' and ``event'' modes collected
X-ray events below and above 13~keV respectively, which were then
merged, accumulated at 4~s intervals, and binned to 25 channels,
approximately logarithmic in energy.  After background subtraction,
the 4~s spectra were statistically significant and also captured most
of the dramatic variability in the source.  Response matrices were
generated by PCARMF v3.5 and corrected for secular gain drift.

Spectra from 3.18--30~keV were fitted individually with an absorbed
two-component emission model consisting of a multicolor ``disk'' black
body (Mitsuda et al \markcite{m&84}1984), representing emission from
an optically thick accretion disk (in steady state), and a power law,
which may be associated with Compton scattering of soft photons by
energetic electrons. A systematic error of 1\%, comparable to the
statistical error, was added in quadrature to account for
uncertainties in the detector response calibration.  The fitted disk
model parameters are the temperature and radius at the inner edge of
the accretion disk.  We chose this model because it is likely to bear
some physical resemblance to the \grs\ system, and because the fits
are acceptable.  In a first pass through the data, neutral absorption
was allowed to vary at each time step; it was noted that the
absorption remained essentially constant at $5.7\times
10^{22}$~\pcm{2}, and in a second pass, it was fixed at that value.
The ensemble of reduced $\chi_{\nu}^2$ values had a mean of 1.48 and
standard deviation of 0.43 for 21 degrees of freedom.  The 1--30~keV
unabsorbed source luminosity varied from $(0.5-3.5)\times
10^{39}$~erg~\psec\ over the observation, assuming a distance of
12.5~kpc.  Figure~\ref{Fcorrel} is a scatter plot of the inner disk
temperature and photon index parameters, and shows that there is a
large degree of ``clustering'' in a few states.  While the absolute
value of the parameters may change with improved calibration, the
relative distribution of values should remain approximately the same.
A preliminary discussion of the results, showing also the parameters
as a function of time, can be found in Swank et al. (1998).

Power density spectra (PDS) from 0.25~to 1024~Hz were also constructed
at 4~s intervals by performing an FFT on the combined data, a portion
of which is shown in the lower panel of Figure~\ref{Flcqpo}.  There is
no evidence that the 67~Hz QPO is manifested in this observation; the
3-$\sigma$ upper limit for the QPO is 0.6\%.  A distinct
variable-frequency QPO does however appear from 2--12~Hz, for example
during the extreme dip between 1100--1600~s in Figure~\ref{Flcqpo}.
In the other observations, the combined frequency range is
approximately 1--15~Hz.  The fractional RMS amplitudes are as high as
10 percent, increasing with energy.  Clearly the frequency of the QPO
depends in some form on the source intensity.  We show below that the
primary dependence is on the disk flux and not the power law flux,
except at the lowest frequencies where the disk flux is very weak.

There are several other features of note in the dynamical power
spectrum: (1) when the source is at its brightest, a strong
low-frequency ($< 5$~Hz) noise component is observed; and (2)
following the ``spike'' at time 1600~s the power spectrum becomes
extremely quiet in terms of variability.  These features persist in
subsequent cycles of the source on the same day (which are not shown),
and show a regular correlation between the source's spectral and
temporal variability behavior.

\section{Spectral and Temporal Correlation \label{Scorr}}
\begin{figure}[tbp] 
\centerline{\epsfxsize=4.3in{\epsfbox{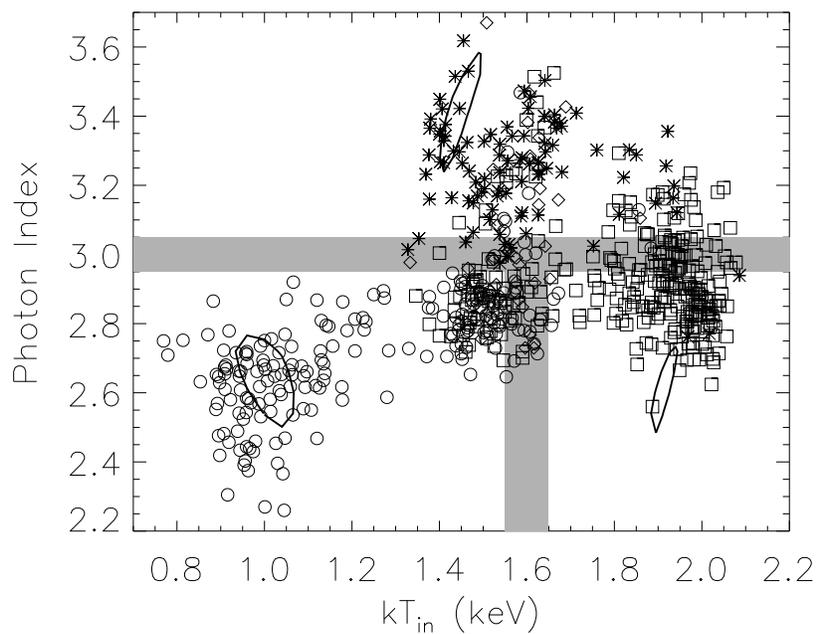}}} 
\caption{ 
Plot of correlation between inner disk temperature, $T_{\rm in}$, and
power law photon index.  The symbols correspond to the temporal
behavior: 1--15~Hz QPO state (circle), low frequency ($< 5$~Hz) noise
state (square), quiet state (star), and an occasional 5~Hz QPO
(diamond).  The grey bars delineate the regions described in
Table~\ref{Tstate}, and the width of the bars represents the
approximate transition region between states.  Representative
confidence contours at a $1\sigma$ level are shown.\label{Fcorrel}}
\end{figure}

Having a record of the properties of both the PDS and energy spectrum,
it is straightforward to compare the two for correlations.  The power
spectra of \grs\ were manually divided into at least three separate
categories, as noted in Figure~\ref{Fcorrel}: a 1--15~Hz QPO state,
a low frequency noise state, and a quiet state.  These temporal states
are sufficiently distinct that they can be visually separated and are
plotted with different symbols in Figure~\ref{Fcorrel}.

It is remarkable that each temporal state occupies a fairly
well-localized domain of the energy spectrum.  The approximate
demarcations between separate states are shown in
Table~\ref{Tstate}. The QPO appears when the inner disk temperature is
relatively low and the power law component is hard.  In addition to
the extreme dip already mentioned, the QPO also appears briefly as
``U''-shaped features at $\sim$8~Hz in the PDS when the spectral
conditions are satisfied, near times 2500~s and 2800~s, as the
magnified region in Figure~\ref{Flcqpo} demonstrates.  Although the
values in Table~1 correspond only to 1997~September~09, in the course
of scanning many observations we find that the spectral/temporal
behavior is similar.  All three states are not always present, but the
separatrix between the states in $kT_{\rm in}$ vs. photon index space
only varies by perhaps as much as $\sim$10\%. The overall
spectral/temporal correlation is thus quite robust.

\begin{figure}[tbp] 
\centerline{\epsfxsize=4.3in{\epsfbox{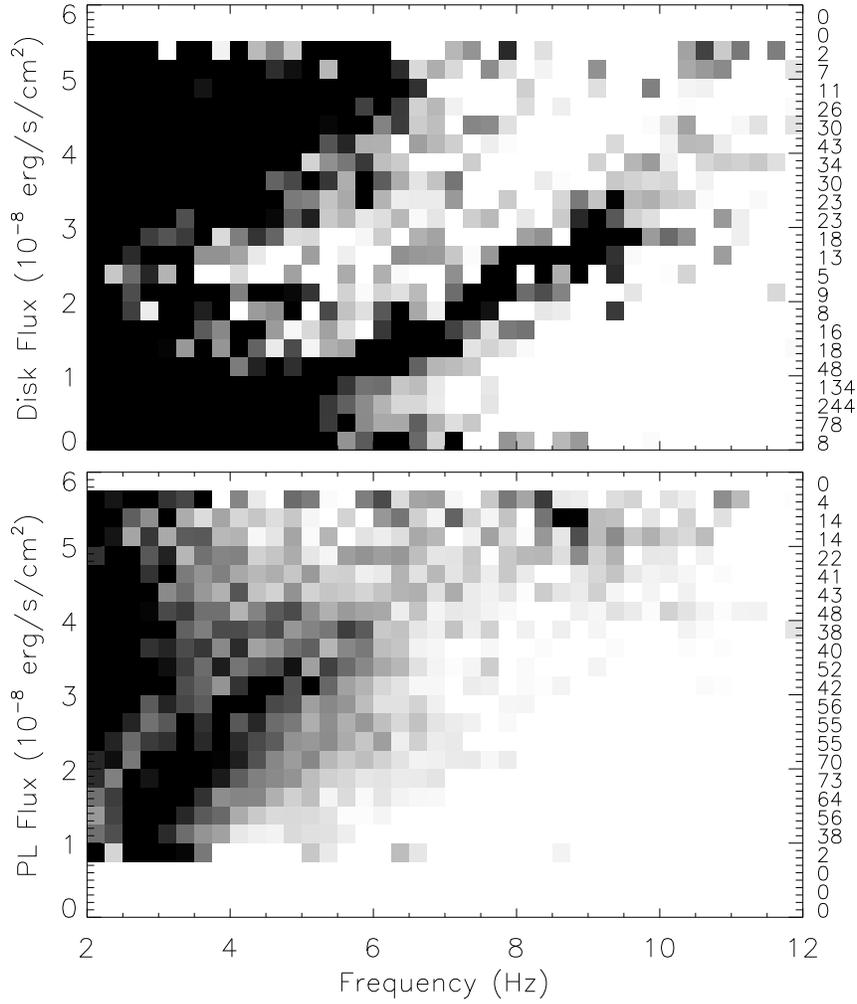}}} 
\caption{ 
A stack of averaged power spectra showing the variation of
the spectral density with black body flux (top), and power law flux
(bottom).  Each row is the average of several individual 4~s power
spectra at the given flux level (the number of averaged spectra is
shown at the right).  The intensity stretches are different, 5--10
(Leahy units; top) and 5--20 (bottom), and designed to accentuate the
dominant correlation in each panel.\label{Fbbcora}}
\end{figure}

The approximate correlation between total intensity and QPO frequency
on time scales of days has been noted previously (CST97, MGR97).
Clearly this general trend holds on the shorter time scales presented
in this Letter, and we wished to determine whether the dependence was
tied distinctly to either the thermal or power law components.  This
was accomplished by constructing an average power spectrum for each
fitted flux level, shown in Figure~\ref{Fbbcora}.  The figure was
constructed by first selecting spectra with a relatively low
temperature and hard power law, using the spectral cuts in
Table~\ref{Tstate}.  By using a spectral cut, we have tried to avoid
the bias introduced by choosing individual spectra by hand.  The
selected spectra are identified with a black mark in the central band
of Figure~\ref{Flcqpo}.  The spectra were then sorted into component
1--30~keV flux levels and averaged.  The result is a power ``image''
representing the average spectral density parameterized by both
frequency and X-ray flux.

The correlation between thermal flux and QPO frequency is very tight
between 4~and 10~Hz, while the correlation with power law intensity is
very weak (except perhaps from 2--4~Hz where the contribution
attributed to the disk is much less than the power law flux, and is
difficult to determine precisely with the PCA).  From a qualitative
standpoint, we could find no other relevant quantity or spectral
parameter, including the total X-ray flux and derived quantities such
as the inner disk radius, that made a tighter correlation than the
thermal flux.  This seems surprising given that the QPOs appear during
the extreme dips when the power law flux is strong.  On the other
hand, the oscillations also appear when the thermal flux is quite
high, for example during the brief episode near 2500~s.  The averaging
technique we have used is sensitive to weaker oscillations that do not
show in a single power spectrum, but are apparent in a flux-selected
average spectrum.

Above 4~Hz the correlation of the frequency with the thermal flux is
quite linear: the frequency grows at $1.70\pm0.05$~Hz for each
increase of $1\times 10^{-8}$~erg~\psec~\pcm{2} in thermal flux.
Below 3--4~Hz the correlation appears to be lost.  In that frequency
range, a rather broad correlation with power law flux is suggested by
Figure~\ref{Fbbcora}.  Upon examination of individual QPO tracks via
peak fitting during the extreme dips, the frequency between 2--4~Hz
does appear to vary strongly with the power law flux, but tracks from
different times do not overlap.  The broad correlation reflects the
variance between QPO tracks.

We emphasize that in examining several observations we find that the
curves of frequency vs. thermal flux correlation overlay one another
quite closely.  The slope and intercept of the linear correlation
changes slightly from observation to observation by perhaps 10
percent.  The tight dependence of QPO frequency on the thermal flux
(and {\it not} the power law flux) thus seems relatively certain. Muno
et al. (1998) undertake to fit the QPO peaks individually, and find
general agreement with our results.

\section{Discussion \label{Sdisc}}

Quasiperiodic oscillations whose frequency depends on flux are seen
from neutron star sources.  The frequencies of both horizontal branch
oscillations (HBO) of accreting ``Z'' sources, and the more recently
discovered kilohertz oscillations, seen in both ``Z'' and ``Atoll''
low-mass X-ray binaries, depend on intensity.  Both the magnetospheric
beat frequency model (Alpar \& Shaham 1985) and the sonic point model
(Miller, Lamb \& Psaltis 1998) are driven by an interaction between
the central rotating compact object and the inner edge of the
accretion disk, and can produce frequency variations.  By analogy, the
variable low frequency oscillations in black hole candidates (CST97;
Takizawa et al 1997) could also be associated with the inner disk
edge.  Indeed, Chen, Taam \& Swank (1998) remark upon the similarity
in frequencies, when scaled by the central mass, and suggest a common
origin.  The beat models, however, rely on radiation or magnetic
fields from the central source, which cannot play a role in black hole
systems.  If the oscillations are at the orbital Kepler frequency of
the disk, they would come from material at a much larger radius than
that of the inner disk: a Kepler radius of 500(3000)~km with 15(1)~Hz
oscillations for a $10M_{\sun}$ black hole, compared with a much
smaller X-ray spectral radius of 60(300)~km (corrected by a factor of
$\sim$3 for scattering effects [Zhang, Cui \& Chen 1997], but not
relativistic effects, which are expected to be $\le 5\%$).
Furthermore, other observations have similar QPO frequencies and disk
fluxes, but substantially different fitted inner radii; this argues
against the interpretation that the oscillations depend solely on the
inner edge of the disk.

The simultaneous presence of a strong, hard power law component during
the 1--15~Hz QPO suggests that the ``corona,'' in whatever form it may
take, plays a vital role in the formation of the oscillations.  The
maximum fractional RMS amplitude of the oscillations increases from
6\% to 13\% in the 2--7 and 15--31~keV bands, respectively, and yet
the thermal disk component cuts off at $\sim 10$~keV.  Nevertheless,
the disk flux is also clearly important, as it appears to regulate the
frequency of the oscillations.  Some properties of the corona may
however be strongly tied to the disk properties.

For a hot coronal inner region, the diffusion time scale could be close
to the periods observed.  The frequency corresponding to the diffusion
time scale works out to be $\propto R^{-1.5}$, where $R$ is the radius.
The ratio of minimum to maximum frequencies (1:15) corresponds very
roughly to the ratio of the radii $(60:300)^{1.5}$ determined by the
spectra. But since the data for even a single observation does not
show a systematic correlation between disk radius and the frequency,
other changes would be needed to modify the radius dependence.

Recent theories of accretion disks find a number of instabilities
which can produce global intensity oscillations in black hole systems.
Simulations of the thermal-viscous instability have led to strong
oscillations from 20--60~mHz, even when the dissipative effects of a
corona are included (Chen \& Taam 1994; Abramowicz, Chen \& Taam
1995).  Taam, Chen \& Swank (1997) have invoked this model to explain
100~s outbursts from \grs.  However, this is a much longer period than
we observe here, and further, predicts a negative rather than positive
correlation between QPO frequency and mass accretion rate (and hence
flux).  Inertial-acoustic modes, sound waves modified by rotation, can
also be excited in the inner disk (e.g., Chen \& Taam 1995).
Simulations have produced oscillations with a few percent amplitudes
at the maximum epicyclic frequency, or $110 (M/10M_{\sun})$~Hz (Chen
\& Taam 1995), but also a 4~Hz modulation corresponding to the
sound-crossing time of the inner unstable portion of the accretion
disk.  While it is conceivable that we are observing this lower
frequency modulation, a mechanism such as scattering must be invoked
to explain the absence of simultaneous higher frequency oscillations.

The oscillation might also be due to oscillations of a shock at the
inner edge of the accretion disk (Molteni, Sponholz \& Chakrabarti
1996).  The predicted oscillation frequency, $f_{\rm QPO} \simeq 11
(M/10M_{\sun})^{-1}$~Hz, is comparable to what we observe.  The
simulated amplitude is several percent, and the frequency increases
with the mass accretion rate.  However, the actual presence of a shock
at the inner edge of the accretion disk is not likely for accretion
flows characterized by high specific angular momentum (Narayan, Kato
\& Honma 1997; Chen, Abramowicz \& Lasota 1997; Chen et al 1997).

In summary, we find that the 1--15~Hz QPOs occur when the hard power
law spectral component becomes strong, especially during extreme dips
when the source is very active. The QPOs themselves have a hard
spectrum which suggests oscillations of the power law component.
However, above 4~Hz we find that the frequency depends nearly linearly
on the flux of the thermal component, and not on the power law
component.  When the thermal flux is lowest, a weak correlation with
the power law flux is suggested.  When the power law flux drops and
the slope steepens at the end of the dips, the QPO then disappears.
This strongly suggests that the two spectral components interact with
each other.  Such considerations have not received much theoretical
treatment in the literature as yet.  A more extensive set of
observations of this phenomenon in \grs\ will be presented by Muno, et
al (1998).

\acknowledgements We appreciate discussions with Ed Morgan, Michael
Muno, Ron Remillard, and David Westbrook.

\newpage 

\newpage

\newpage
\begin{table}[tbp]
\caption{Spectral/Temporal States of \grs\label{Tstate}}
\begin{tabular}{lcc}
\hline
Temporal State & $kT_{\rm in}$& Photon Index \\
\tableline
1--15~Hz QPO & $< 1.55$~keV  & $< 2.95$        \\
Low Freq. Noise& $> 1.65$~keV  & $< 2.95$        \\
Quiet          & ---\tablenotemark{a}  & $> 3.05$        \\
\end{tabular}
\tablenotetext{a}{all temperatures}
\end{table}

\end{document}